\documentclass[prb,twocolumn,superscriptaddress,showpacs,amsmath,amssymb]{revtex4}
\usepackage{color}
\usepackage{bm}% bold math

\usepackage{graphicx}
\begin{document}

\title{Universal mechanism of dissipation in Fermi superfluids at ultra low temperatures}

\author{Mihail A. Silaev}
 \affiliation{Institute for Physics of Microstructures RAS, 603950
Nizhny Novgorod, Russia.}
 \affiliation{Department of Theoretical
Physics, The Royal Institute of Technology, Stockholm, SE-10691
Sweden}

\date{\today}

\begin{abstract}
We show that the vortex dynamics in Fermi superfluids at ultra-low
temperatures is governed by the local heating of the vortex cores
creating the heat flux carried by non-equilibrium quasiparticles
emitted by moving vortices. This mechanism provides a universal
zero temperature limit of dissipation in Fermi superfluids. For
the typical experimental conditions realized by the turbulent
motion of $^3$He-B the temperature of vortex cores is estimated to
be of the order $0.2 T_c$. The dispersion of Kelvin waves is
derived and the heat flow generated by Kelvin cascade is shown to
have the value close to the experimentally observed.
\end{abstract}
\maketitle

Vortex dynamics determines the most fundamental properties of
supefluids and superconductors. The well known textbook example is
the motion of Abrikosov vortices under the action of the Lorentz
force which determine the finite resistance of the type-II
superconductors\cite{SaintJames}. Equally important is the vortex
dynamics in the superfluids $^4$He and $^3$He where motion of
quantized vortices mediate the mutual friction force between the
normal and superfluid components originating from the scattering
of normal excitations by moving vortex lines. The dissipative
component of mutual friction provides the relaxation of superflow
which is the key aspect in the theory of superfluid
turbulence\cite{VinenIntroduction}. Recently there has been a
renewal interest in this field  owning to the developing
experimental techniques allowing to study in particular the decay
of quantum turbulence at very low temperatures
\cite{EltsovReveiw}. In this regime
  the normal component of the fluid becomes
 extinct and the mutual friction can not provide the energy
 dissipation and the relaxation of the superflow. The state-of-art experiments
 \cite{Bradley,Eltsov,Walmsley}
 demonstrate that the dissipation in the vortex motion remains
 finite at the lowest temperatures both in the Bose superfluid $^4$He
 and Fermi one $^3$He-B. The energy dissipated by turbulence is
 supposed to be released in the form of the thermal flux of non-equilibrium quasiparticles
 which recently has become accessible to direct
 measurements\cite{BradleyNature,EltsovHeatFlux}.
 These experimental results pose a challenging
 question about the fundamental nature of the dissipation in
 superfluids at ultra low and even zero temperatures.

In this Letter we propose the new mechanism which governs the
vortex dynamics in Fermi superfluids at ultra-low temperatures. We
show that even at zero temperature in the absence of the normal
component the non-stationary dynamics of vortices is intrinsically
dissipative as a result of relaxation processes inside the vortex
core moving with the finite acceleration (which is a generic
situation in superfluid turbulence).
 We address the case of typical superconductors and superfluid Fermi systems
 which can be described within the weak coupling Bardeen-Cooper-Shcriffer (BCS) theory.
 In particular such restrictions means that the critical temperature and energy gap are much less
than the Fermi energy which allows the so-called quasiclassical
description of the pairing state\cite{KopninReview}. Note that for
the case of cold atomic gases this description is valid only in
deep BCS limit.

  %%%%%%%%%%%%%%%%%%%%%%%%%%%%%%%%%%%%%%%%%%%%%%%%%%%%%%%%%%
  \begin{figure}[!h]
 \centerline{\includegraphics[width=1.0\linewidth]{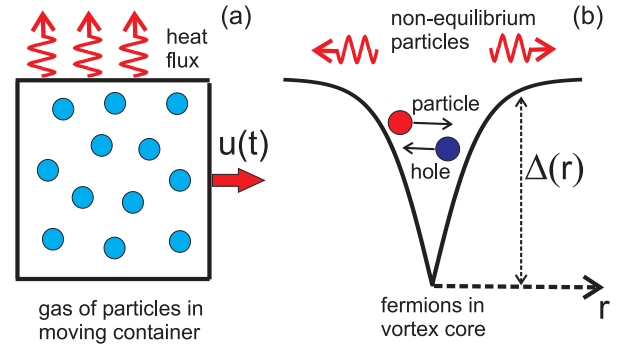}}
 \caption{\label{Fig1} (a) Container with the ideal gas of
 particles demonstrating the classical analog of vortex core. The
 temperature of gas is raised in case if ${\bf \dot{u}}\neq 0$.
 The energy dissipation by the heat flux through the heat conducting wall is shown
 schematically. (b) Sketch of the vortex core and
 quasiparticle confinement due to Andreev reflection and the flow
 of non-equilibrium delocalized quasiparticles.}
   \end{figure}
%%%%%%%%%%%%%%%%%%%%%%%%%%%%%%%%%%%%%%%%%%%%%%%%%%%%%%%%%%%%%%%%%

The mechanism of the ultra-low temperature energy dissipation in
Fermi superfluids has a very transparent classical physics
analogy. As will be discussed below the vortex core can be viewed
as an insulating box containing a gas of monoatomic molecules
which represent localized vortex core quasiparticles shown in
Fig.\ref{Fig1}(a). Let us assume that the oscillating center of
mass motion of the box is produced characterized by the velocity
$u (t) =u_0 \Theta (t)$, where $\Theta (t)$ is a step function
with a period $t_p$ larger than the thermalization time of the
gas. Then each step change of the velocity will result in the
temperature rise which can be calculated from the energy
conservation law to be $\Delta T_{loc}= 4u_0^2/c_v$ where $c_v$ is
the specific heat per unit mass. Averaging over the time interval
much larger than the period we obtain the linear temperature
growth in time $T_{loc} (t)= \frac{4u_0^2}{c_v}\frac{t}{t_p}$.
Furthermore, the temperature growth will certainly stop at some
point regulated by the finite heat conductivity of the container
walls. The balance of the heating by box acceleration and the heat
release through the heat conducting walls will set the stationary
value of the temperature $T_{loc}$ of the particles localized
inside the box. The heat flow through the container walls
determines the rate of the energy dissipation in the system.

At temperatures considerably lower than $T_c$ vortex dynamics in
clean type-II superconductors and Fermi superfluids is determined
by the kinetics of localized excitations bound in the vortex cores
driven out of equilibrium by vortex motion \cite{KopninReview}. In
Fermi superfluid containing vortices the profile of the order
parameter $\Delta \left( {\bf r} \right) $ near the vortex core
produces a potential well where localized states with a discrete
spectrum exist. The quasiparticle confinement results from the
subsequent Andreev reflections transforming particles to the holes
and vice versa [Fig.\ref{Fig1}(b)]. The localized states
correspond to energies $|\epsilon_{loc} |<\Delta _0$ where $\Delta
_0$ is bulk value of the gap. The spectrum has the so called
anomalous branch\cite{Caroli} whose energy varies from $-\Delta
_0$ to $+\Delta_0$ and crosses $\epsilon =0$. The anomalous energy
branch has topological origin \cite{VolovikBook} and exists for
all kinds of quantized vortices in the weak coupling Fermi
superfluids which allows for the quasiclassical description. In
model axially symmetric singly-quantized vortices the anomalous
branch states are characterized by the particle angular momentum
projection on the vortex axis $\mu$. For low energies
$|\epsilon_{loc}| \ll \Delta _{0}$, the anomalous branch is
$\epsilon _{loc}=-\hbar \omega _{0}\mu $. The separation between
the levels with neighboring angular momenta $\mu$ and $\mu+1$ is
$\omega_0\sim \Delta_0/(\hbar k_F\xi)$ where $k_F$ is the Fermi
momentum and $\xi$ is the coherence length. In the weak coupling
limit the interlevel distance is small $\omega_0\ll \Delta_0$
therefore the angular momentum $\mu$ can be considered as
classical variable commuting with the corresponding angle
$\theta_p$, characterizing the direction of quasiparticle motion
in plane perpendicular to the vortex axis.

Our results remains valid with  some modifications for the
non-axisymmetric vortices existing in multicimponent Fermi
superfluids like $^3$He \cite{VolovikRMP}. In this case one should
assume the angular dependence of bound states energy
 $\epsilon _{loc}=\epsilon_{loc}(\mu,\theta_p)$\cite{Stone,VolovikBook}.
 Note also that in general the singly-quantized vortices have two anomalous energy branches
  corresponding to different spin states of bound
  fermions\cite{SilaevHe3}. The kinetic theory that we will
  use allows to treat each of the anomalous branches in separate.
  Therefore below we omit the spin indices, which finally reduces the full
Bogoliubov-de Gennes $4\times 4$ matrix theory to the effective
scalar theory of the core fermions.

The key idea of the present work is that the ensemble of
quasiparticles localized within the moving vortex core is in some
sense analogous to the particle gas confined in a moving box.
Below we will show that the local heating of the localized
quasiparticles due to the vortex acceleration can lead to the
temperature growth which is compensated by the heat flow out of
the vortex cores in the form of the non-equilibrium delocalized
quasiparticles.

The equilibrium distribution function of Fermi quasiparticles has
the form $f^{(0)}=\tanh(\epsilon/2T)$. The delocalized
quasiparticles with energies $|\varepsilon|>\Delta_0$ have the
temperature $T=T_{del}$ fixed by the heat bath. We assume that
$T_{del}\rightarrow 0$ which allows to neglect the concentration
of delocalized particles so that
$f_{del}^{(0)}(\epsilon>\Delta_0)=1$ and
$f_{del}^{(0)}(\epsilon<-\Delta_0)=-1$. On the other hand as we
will see below the localized particles have the different
temperature $T=T_{loc}$ which can be much larger $T_{loc}\gg
T_{del}$.

We start with a kinetic equation\cite{KopninReview} for the
distribution function of localized quasiparticles
$f=f(t,\mu,\theta_p)$ which depends on time $t$ and a pair of
canonical variables $(\mu,\theta_p)$
\begin{equation}\label{kinetic}
 -\omega_0\left([{\bf p}\times {\bf u}]\cdot {\bf z}\right) \frac{\partial f}{\partial
\varepsilon} + \frac{\partial f}{\partial t} + \frac{\partial
f}{\partial
 \theta_p}\dot{\theta_p}+ \frac{\partial f}{\partial
 \mu}\dot{\mu}=St(f).
\end{equation}
Here the energy spectrum of localized particles
$\epsilon_{loc}=-\hbar\omega_0\mu$ plays the role of the classical
Hamiltonian governing the dynamics of the canonical variables so
that $\dot{\theta}_p=-\omega_0$ and $\dot{\mu}=0$. The source of
non-equilibrium is the first term in Eq. (\ref{kinetic})
 which is generated by the vortex motion with
the velocity ${\bf u}=(u_x{\bf x},u_y{\bf y})$.

 The collision integral (CI) in the
r.h.s. of this equation (\ref{kinetic}) can be determined by
different scattering processes which depend on the particular
system where the Fermi superfluidity is considered. We will
consider the universal mechanism which inherently exists in any
Fermi liquid - the mutual scattering of quasiparticles. In $^3$He
and ultra-cold Fermi gases the collision integral is determined
ultimately by this mechanism.

The general form of the particle-particle CI has rather
complicated form \cite{Eliashberg}. Therefore we use the model
forms of the CI which features however the important properties of
the exact CI. We are interested in the relaxation of localized
quasiparticle with the energy $|\epsilon|<\Delta_0$ and momentum
${\bf p}$. It collides with a particle characterized by
$\epsilon_1,{\bf p_1}$ and they scatter into states
$\epsilon_2,{\bf p_2}$ and $\epsilon_3,{\bf p_3}$. During this
process the energy and momentum are conserved
$\epsilon+\epsilon_1=\epsilon_2+\epsilon_3$ and ${\bf p}+{\bf
p}_1={\bf p}_2+{\bf p}_3$. Due to the Gallilean invariance the CI
integral vanishes when all quasiparticles are moving with a net
velocity ${\bf w}$ having the equilibrium distribution function
$f^{(0)}=\tanh[(\epsilon-{\bf p w})/2T]$.

In general, there are three types of scattering processes which
contribute to the particle- particle CI which we consider in
separate.

{\bf (i)} Collision of localized and delocalized qusiparticles
when $|\epsilon|,|\epsilon_1|<\Delta_0$ and at the same time
  $|\epsilon_{2,3}|>\Delta_0$. This is the basic scattering process
determining the force acting on the vortex from the heat
bath\cite{Kopnin2times}.  For $T_{del}\rightarrow 0$ the
probability of this collision is negligibly small and for
$T_{del}=0$ the exact calculation
  shows that the corresponding part of the CI is zero.

 {\bf (ii)} The scattering involving only localized quasiparticles
  $|\epsilon|, |\epsilon_1|,|\epsilon_2|, |\epsilon_3|<\Delta_0$.
 In this case the collision time approximation can be used with
 account of the Gallilean invariance in the system of localized
 particles \cite{Kopnin2times}
 \begin{equation}\label{Eq:CIlocalized}
 St_{loc}(f)=\frac{1}{\tau_{loc}}[f^{(0)}(\epsilon-{\bf
 pw})-f].
 \end{equation}
 Here ${\bf w}$ is the 'drag velocity'
 which determines the evolution of distribution function. In the
 absence of collisions with delocalized states it coincides with
 the vortex velocity ${\bf w}={\bf u}$. The collision time depends on the temperature
 of localized quasiparticles and for $T_{loc}\ll \Delta_0$ was estimated to be  \cite{Kopnin2times}
  $\tau^{-1}_{loc}\sim \tau^{-1}_{n} (T_{loc}/\Delta_0)^4$ where
  $\tau_n^{-1}\sim \Delta_0^2/E_F$ is the normal state scattering
  rate at $T=T_c$. Importantly this part of CI involving only localized particles
  (\ref{Eq:CIlocalized}) conserves the energy of localized
  particles
 \begin{equation}\label{Eq:EnergyConservation}
   \int_{-\Delta_0}^{\Delta_0} \epsilon St_{loc} (f) d\epsilon =0.
\end{equation}

{\bf (iii)} Collision of two localized quasiparticles with
production of localized and delocalized ones. In this case we
assume that $|\epsilon|, |\epsilon_1|,|\epsilon_2| <\Delta_0$ and
$|\epsilon_3|>\Delta_0$. During such processes the non-equilibrium
delocalized quasiparticles are created which carry the energy flow
out of the vortex cores shown schematically in Fig.\ref{Fig1}(b).
This mechanism of energy losses is similar to the heat flow
through the heat conducting wall in the gas container  discussed
above [see Fig.\ref{Fig1}(a)].

To calculate the probability of scattering event {\bf (iii)} one
needs to take into account the contributions from states with
$\epsilon_3>\Delta_0$ and $\epsilon_3<-\Delta_0$. We will consider
in detail only the first case and the second one can be treated
analogously. We can put $f^{(0)}(\epsilon_3)=1$ which means that
the final delocalized state is empty. The scattering probability
is determined by the factor
$F(\epsilon,\epsilon_1,\epsilon_2)=[1-f(\epsilon)][1-f(\epsilon_1)][1+f(\epsilon_2)]$
which is the product of the initial states occupational numbers
and the probability of the final localized state to be empty. To
obtain the total scattering probability we should integrate over
the energies $\epsilon_{1,2}$ taking into account the energy
conservation requirement $-\Delta_0<\epsilon_2<\epsilon^*$ where
$\epsilon^*=\epsilon_1+\epsilon-\Delta_0$ which yields in
particular $\Delta_0>\epsilon_1>-\epsilon$. Then we obtain the
following contribution to CI
\begin{equation}\label{Eq:CIloc-del}
 St_{l-dl}(f)=(\tau_n \Delta_0^2)^{-1}\int_{-\epsilon}^{\Delta_0} d\epsilon_1
\int_{-\Delta_0}^{\epsilon^*} d\epsilon_2
F(\epsilon,\epsilon_1,\epsilon_2)
\end{equation}
where the prefactor can be determined from exact form of the CI.
The contribution from the states with $\epsilon_3<-\Delta_0$
yields the same result Eq.(\ref{Eq:CIloc-del}) up to the numerical
prefactor which depends on the particle-hall asymmetry. Note that
in contrast to the Eq.(\ref{Eq:EnergyConservation}) the part of
the CI (\ref{Eq:CIloc-del}) does not conserve the energy of
localized quasiparticles and therefore provides the energy flow
out of the vortex cores.

Having in hand the expressions for the CI we can find the
corrections to the distribution functions generated by the vortex
motion. At first we find the anisotropic corrections taking into
account only the localized part of the CI (\ref{Eq:CIlocalized})
$St(f) =St_{loc} (f)$. Our key idea is that to induce the
non-equilibrium distribution in Gallilean  invariant system of
localized quasiparticles the accelerated motion of the vortex is
needed $\dot{{\bf u}}\neq 0$. The solution of the kinetic
equations  can be found in the usual way. Substituting the ansatz
 $ f= f^{(0)}(\epsilon)- ({\bf pu})\frac{\partial
 f^{(0)}}{\partial\epsilon}+f^{(1)}$ to the Eq.(\ref{kinetic}) we obtain
 the expression for the non-equilibrium correction:
\begin{equation}\label{Eq:Solution1}
f^{(1)}=\tau_{loc} \left( \gamma_{\parallel}{\bf p\cdot \dot{u}} +
\gamma_{\perp} [{\bf \bf p \times \dot{u}}]\cdot{\bf z_v} \right)
\frac{\partial f^{(0)}_{loc}}{\partial \varepsilon}
\end{equation}
where ${\bf z_v}$ is a unit vector determined by the direction of
vorticity, $\gamma_{\parallel}= 1/[1+(\omega_0\tau_{loc})^2]$ and
$\gamma_{\perp}=\omega_0\tau_{loc}\gamma_{\parallel}$.

The force acting on the moving vortex from the ensemble of
localized quasiparticles can be calculated substituting the
solution (\ref{Eq:Solution1}) to the conventional
expression\cite{KopninReview}:
 \begin{equation}\label{Eq:Force1}
  {\bf F_{loc}}= -M_\parallel {\bf \dot{u}}-M_\perp [{\bf z_v}\times{\bf
 \dot{u}}] \ ,
 \end{equation}
 where the masses are determined by
  \begin{equation}\label{Eq:Mass}
  M_{\parallel (\perp)}=
 \frac{\pi\hbar N}{2} \left\langle
 \int_{-\Delta_0}^{\Delta_0} \tau_{loc}\gamma_{\parallel (\perp)}
 \frac{\partial f^{0}_{loc}}{\partial \varepsilon}
  d\varepsilon\right\rangle_{F.S.}
\end{equation} where $N$ is the particle concentration and
the brackets denote the Fermi surface averaging. The first term in
the Eq.(\ref{Eq:Force1}) is determined  by the inertial vortex
mass $M_\parallel$ which in the limit $\tau_{loc}=\infty$
coincides with the expression obtained in collisionless regime
\cite{KopninReview}. This inertial term does not lead to the
dissipation, that is to the transformation of the vortex kinetic
energy to the heat. On the contrary the second term in the
Eq.(\ref{Eq:Force1}) can provide the dissipation. It is easy to
see that for the circular vortex motion the period-averaged work
of force ${\bf F_{loc}}$ is non-zero due to the second term. In
case of non-axially symmetric vortices the expressions for masses
$M_{\parallel (\perp)}$ become more involved and take into account
the interplay between bound fermions and the modes of rotational
vortex core motion\cite{RotatingVortex}.

 Let us consider the generic
example of the single vortex line motion in the self-induced
superfluid velocity field described in terms of the propagation of
Kelvin waves (KW) along the vortex line. KW play a key part in the
relaxation of superfluid turbulence at low temperatures. The KW
cascade characterized by the universal spectrum\cite{KS04,Lvov}
carries the turbulent energy  from the main length scale of the
vortex line separation to essentially lower length scales.

In the local induction approximation (LIA) valid for the large
Kelvin wavelength the motion of vortex line\cite{Sonin} is
determined by the linear tension energy, Magnus force $F_M=-\kappa
N [{\bf z_v\times u}]$ where $\kappa=2\pi\hbar/m$ is a vorticity
and the force (\ref{Eq:Force1}) exerted on the vortex line by the
localized excitations
\begin{equation}\label{Eq:KWmotion}
m_\parallel {\bf \ddot{\bf U}}+m_\perp [{\bf z_v}\times{\bf
 \ddot{{\bf U}}}] +[{\bf z_v}\times \dot{{\bf U}}]=\nu_s {\bf U}_{zz}
\end{equation}
  where ${\bf U} (z,t)$ is the vortex displacement,
   $\nu_s=\kappa \Lambda/4\pi$ and $\Lambda=\ln (\xi/l_c)$ where $l_c$ is the cutoff
   parameter, $m_{\parallel,\bot}=M_{\parallel,\bot}/(\pi N)$. The
dispersion relation following from the Eq.(\ref{Eq:KWmotion}) has
the form
 \begin{equation}\label{Eq:KWdispersion}
 \omega_K= \pm k^2\nu_s+(k^2\nu_s)^2(im_\perp-m_\parallel).
 \end{equation}
 The above expression can be extrapolated to the smaller
 wavelengths\cite{Sonin} by setting the cutoff parameter
 $l_c=1/k$. Note that the corrections to the KW frequency of the
 same type as given by Eq.(\ref{Eq:KWdispersion}) are
 necessarily taken into account to break the integrability of LIA and launch the KW
 cascade (see e.g. Ref.\onlinecite{KS04,LvovBottleneck}).
 Thus the corrections given by Eq.(\ref{Eq:KWdispersion}) can in principle have significant influence
 on the theory of quantum turbulence.

 The expression (\ref{Eq:KWdispersion})
 demonstrates that even in the absence of the heat bath the
 non-trivial dynamics of localized excitation can provide the
 finite decrement which can be estimated as $\tau_K^{-1}=m_\perp\omega_K^2\sim
\tau_{loc}^{-1}\left(\omega_K/\omega_0\right)^2$.
 The decrement depends on the temperature $T_{loc}$
 and can be finite even if the heat bath temperature is zero
 $T_{del}=0$.

To calculate the temperature $T_{loc}$ we use the method analogous
to the one employed in\cite{KopninUltraLow} with the only
difference that the energy flow out of the vortex core is
determined by the collisions described by the CI
(\ref{Eq:CIloc-del}) and not by the collisions with phonons. Let
us substitute the distribution function (\ref{Eq:Solution1}) into
the Eq. (\ref{kinetic}) and take the average by quasiparticle
momentum direction over the Fermi surface and time $t$. From the
first term in the Eq.(\ref{kinetic}) we obtain
$$
  \left\langle \omega_0[{\bf p\times{u}}]\cdot{\bf z} \frac{\partial f^{(1)}}{\partial
 \epsilon}\right\rangle_{t, FS}=
 \alpha\bar{\gamma}_\perp  \frac{d^2f^{(0)}}{d\epsilon^2}  ,
 $$
 where $\alpha=\left\langle u_x
 \dot{u}_y-u_y \dot{u}_x \right\rangle_t/2$ and
$\bar{\gamma}_\perp=\langle \gamma_\perp \rangle _{FS}$. We have
neglected the term containing the time average $\left\langle
\dot{u}^2\right \rangle_t$ assuming the periodic vortex motion.
Also we multiply and integrate the Eq. (\ref{kinetic}) by energy.
  Making use of the energy conserving property of localized CI (\ref{Eq:EnergyConservation}) we
obtain the equation which determines the temperature $T_{loc}$
 \begin{equation}\label{Eq:Temperature0}
    \int_{-\Delta_0}^{\Delta_0}
   \epsilon St_{l-dl} (f) d\epsilon=\alpha\bar{\gamma}_\perp
  \int_{-\Delta_0}^{\Delta_0} \epsilon
   \frac{d^2f_0}{d\epsilon^2}d\epsilon
\end{equation}
 The obtained equation has the simple physical meaning expressing
  the balance between the heat produced by the accelerated vortex
  motion and the energy flow out of the vortex core carried by the
  emitted delocalized quasiparticles.

 The further analytical analysis can by carried out assuming that
 $T_{loc}\ll \Delta_0$. By the order of magnitude we have $\omega_0 \sim \tau_n^{-1} \sim \Delta_0/(k_F\xi)$
 therefore $\omega_0\tau_{loc}\ll 1$ and $\bar{\gamma}_\perp\sim
 \omega_0\tau_{loc}$. Taking into account the temperature
 dependence of the collision time $\tau_{loc}$ we obtain the
 simplified heat balance equation in the form
  \begin{equation}\label{Eq:Temperature}
  \alpha \xi/V_c^3= \left(\Delta_0/T_{loc}\right)
  e^{-\Delta_0/T_{loc}},
 \end{equation}
 where $V_c=\Delta_0/p_F$ is the critical superfluid velocity.

 To estimate the value of the vortex core temperature $T_{loc}$
 which can be realized in the conditions of quantum turbulence in $^3$He-B
 we evaluate the
 factor $\alpha$ in (\ref{Eq:Temperature}) assuming that the
 vortex dynamics is regulated by the KW cascade. Introducing the complex field
 $W=U_x+iU_y$ we obtain $ \alpha=\left\langle i \dot{W} \ddot{W}^* +
 c.c\right\rangle_t/4 $. The time average can be expressed through the
 'kelvon' occupational number $n_k=L\langle|W_k|^2\rangle$ where
 $W=\kappa^{-1/2} \sum_k W_k (t) e^{ikz}$ and $L\sim l^{-2}$ is the vortex
 line length per unit volume, $l$ is the vortex line separation.
 To obtain the
  estimation of $\alpha$ we can use the spectrum of
  kelvons
  $ n_k=\kappa k^{-3} (k l)^{-\gamma}$
where $\gamma=2/5$ or $\gamma=2/3$ according to Refs.
\onlinecite{KS04} and \onlinecite{Lvov} correspondingly. We have
omitted the dimensionless prefactors which are not important in
our derivation. The above scaling law results in the estimation
$\alpha\sim \kappa^3k_c^4 (k_c l)^{-\gamma}$
% \begin{equation}\label{Eq:alpha2}
%   \alpha\sim \kappa^3k_c^4 (k_c l)^{-\gamma}
%\end{equation}
where $k_c$ is the cutoff wave number of the KW cascade
\cite{VinenCutOff}. The heat balance equation
(\ref{Eq:Temperature}) can be rewritten in the form
 \begin{equation}\label{Eq:Temperature2}
    \left(\Delta_0/T_{loc}\right)
  e^{-\Delta_0/T_{loc}}\approx (k_c\xi)^{4-\gamma}
  (\xi/l)^{\gamma}.
\end{equation}

Further we need to estimate the value of the cutoff wave number
$k_c$ of the KW cascade. This can be obtained in the usual way by
comparing the large-scale energy flow in a Kolmogorov
cascade\cite{VinenCutOff} $E_t\sim \kappa^3l^{-4}$ to the value of
the energy dissipated by the KW cascade $P_{kw}=l^{-2}\int^{k_c}
\tau_K^{-1}\varepsilon_K dk$ where $\varepsilon_K=\omega_k n_k$.
From the equation $P_{kw}=E_t$ we obtain the cutoff parameter
\begin{equation}\label{Eq:kc}
 k_c\sim \xi^{-1}
 \left(\xi/l\right)^{\beta}\left(\Delta_0/T_{loc}\right)^{2\beta}
\end{equation}
where $\beta=(4-\gamma)^{-1}$. Substituting the obtained value of
$k_c$ to the (\ref{Eq:Temperature2}) and taking into account that
for the typical experiments in $^3$He-B $\xi/l\sim 10^{-4}$ we
obtain the temperature $T_{loc}\approx 0.1 \Delta_0\approx 0.2
T_c$. The value of the cutoff parameter is $k_c\sim 0.2 \xi^{-1}$
corresponding to the Kelvin wavelength $\lambda=2\pi/k_c\sim 30
\xi$ which is much larger than the core size $\xi$ and therefore
can be reached by the KW cascade. Thus we obtain that at
temperatures lower than $0.2 T_c$ the dynamics of the vortices in
$^3$He-B should be determined by the local heating of the vortex
cores rather than by the interaction with the heat bath.

It is interesting also to compare the power of the energy losses
 predicted by the theory with the experimentally observed values of the energy
 flux out of the turbulent region generated by the propagating
 vortex front in $^3$He-B at the temperatures down to $0.2 T_c$ \cite{EltsovHeatFlux}.
 We use the typical $^3$He parameters
 $\xi=10^{-6} cm$, $(k_F\xi)\sim  10^2$, $\Delta_0\sim  10^{-25}
 J$,$\tau_n \sim  10^{-6} s $ and assume $T_{loc}\approx 0.1 \Delta_0$
to obtain that $ P_{kw}\sim 10^{-10}(k_c\xi)^{4-\gamma} W/cm^3$.
In the experiment the typical value of the turbulent region is
$\sim 1 cm$ the heat flux is $I_E\approx
10^{-10}(k_c\xi)^{4-\gamma} W/cm^2$
 which is close to the experimentally observed value if $(k_c\xi)\sim 1$.
 Note that this estimation is the minimal value of the heat flux
 which can be obtained in the limit $T\rightarrow 0$.

 In conclusion we have studied the dynamics of quantized vortices
 in ultra cold Fermi superfluids. The new mechanism is suggested of how the energy
 can be transferred to the ensemble of localized quasipartices
 resulting in the local temperature rise inside the vortex cores and dissipation due to the
 escape of hot quasiparticles into delocalized states.
 The force acting on vortex is derived and shown to provide the finite
 decrement of Kelvin waves which is determined by the vortex core temperature.
  It would be interesting to investigate
 the dissipative vortex dynamics under the action of this force in more
 detail in large-scale numerical simulations.

The work is supported by the 'Dynasty' Foundation, Presidential
RSS Council (Grant No. MK-4211.2011.2) and Russian Foundation for
Basic Research. Discussions with G.E. Volovik, N.B. Kopnin, A.S.
Mel'nikov, V.B. Eltsov and  M. Krusius are greatly acknowledged.
%%%%%%%%%%%%%%%%%%%%%%%%%%%%%%%%%%%%%%%%%%%%%%%%%%%%%%%%

\end{document}